\documentstyle[preprint,aps,floats]{revtex}

\begin{document}

\draft

\preprint{\rightline{ANL-HEP-PR-96-28}}

\title{QUARKONIUM DECAY MATRIX ELEMENTS FROM QUENCHED LATTICE QCD}

\author{G.~T.~Bodwin and D.~K.~Sinclair}
\address{HEP Division,
Argonne National Laboratory, 9700 South Cass Avenue, Argonne, IL 60439}
\author{S.~Kim}
\address{Center for Theoretical Physics,
Seoul National University, Seoul, Korea} 

\maketitle

\begin{abstract}
We calculate the long-distance matrix elements for the decays of the
lowest-lying S- and P-wave states of charmonium and bottomonium in quenched
lattice QCD, using a nonrelativistic formulation for the heavy quarks. (The
short-distance coefficients are known from perturbation theory.) In particular,
we present the first calculation from QCD first principles of the color-octet
contribution to P-wave decay---a contribution that is absent in potential
models. We also give the relations between the lattice matrix elements and
their continuum counterparts through one-loop order in perturbation theory. 
\end{abstract}


\setcounter{page}{1}
\pagestyle{plain}
\parskip 5pt
\parindent 0.5in

Heavy quarkonium systems (charmonium, bottomonium) are nonrelativistic: 
in the CM frame, the average quark velocity $v$ satisfies $v^2
\ll 1$ ($v^2 \approx 0.3$ for charmonium, and $v^2 \approx 0.1$ for
bottomonium). Bodwin, Braaten and Lepage (BBL) \cite{bbl} have shown,
within the framework of nonrelativistic QCD (NRQCD), that the smallness
of $v^2$ allows one to express a quarkonium decay rate as a sum of terms,
each of which consists of a long-distance (distance $\sim 1/M_Q v$)
matrix element of a 4-fermion operator in the quarkonium state
multiplied by a short-distance (distance $\sim 1/M_Q$) parton-level
decay rate, which may be calculated perturbatively. In particular, the
decay rates for S-wave quarkonium through next-to-leading order in $v^2$
are given by 
\begin{eqnarray}
\Gamma(^{2s+1}S_J\rightarrow X) & = & {\cal G}_1(^{2s+1}S_J)\,2\,
					{\rm Im\,}f_1
                                      (^{2s+1}S_J)/M_Q^2 
                                      \nonumber \\
                                & + & {\cal F}_1(S)\,2\,{\rm Im\,}g_1
                                      (^{2s+1}S_J)/M_Q^4.
\end{eqnarray}
To the lowest non-trivial order in $v^2$, the 
P-wave decay rate is given by 
\begin{eqnarray}
\Gamma(^{2s+1}P_J\rightarrow X) & = & {\cal H}_1(P)\,2\,{\rm Im\,}f_1
                                      (^{2s+1}P_J)/M_Q^4
                                      \nonumber \\
                                & + & {\cal H}_8(P)\,2\,{\rm Im\,}f_8
                                      (^{2s+1}S_J)/M_Q^2.
\end{eqnarray}
The $f$'s and $g$'s are proportional to the
short-distance rates for the annihilation of a $Q\bar Q$ pair from the 
indicated $^{2s+1}L_J$ state, while ${\cal G}_1$, ${\cal
F}_1$, ${\cal H}_1$, and ${\cal H}_8$ are the long-distance matrix
elements. The subscripts $1$ and $8$ indicate whether the $Q\bar Q$ pair 
is in a relative color-singlet or color-octet state.
In this paper, we report a lattice calculation of the
long-distance matrix elements in QCD for the
lowest-lying S- and P-wave charmonium and bottomonium states.  The 
calculation of ${\cal H}_8$ yields the first result for a heavy-quark
color-octet matrix element that is based on QCD first principles. 

The long-distance matrix elements are defined by
\begin{mathletters}
\label{matrix-elements}
\begin{eqnarray}
{\cal G}_1 & = & \langle ^1S|\psi^{\dag}\chi\chi^{\dag}\psi|^1S\rangle 
\label{eqn:g1}, \\
{\cal F}_1 & = & \langle ^1S|\psi^{\dag}\chi\psi^{\dag} 
(\frac{-i}{2}\stackrel{\leftrightarrow}{\bf D})^2\chi |^1S\rangle, \\
{\cal H}_1 & = &\langle ^1P|\psi^{\dag}(i/2)\stackrel{\leftrightarrow}{\bf D}
\chi.\chi^{\dag}(i/2)\stackrel{\leftrightarrow}{\bf D}\psi|^1P\rangle, \\
{\cal H}_8 & = & \langle ^1P|\psi^{\dag}T^a\chi\chi^{\dag}T^a\psi|^1P\rangle.
                                                               \label{eqn:h8}
\end{eqnarray}                                                   
\end{mathletters}%

The terms proportional to ${\cal G}_1$ and ${\cal H}_1$ in the decay
rates are those that appear in the conventional, color-singlet model
\cite{color-singlet-model}. In the vacuum-saturation approximation
\cite{bbl}, which is correct up to terms of order $v^4$, ${\cal G}_1 =
\frac{3}{2\pi}|R_S(0)|^2$ and ${\cal H}_1 = \frac{9}{2\pi}|R'_P(0)|^2$,
where $R(0)$ is the radial wavefunction at the origin and $R'(0)$ is the
derivative of the radial wavefunction at the origin. The matrix-element
forms of (\ref{matrix-elements}) serve to define a regularized $R(0)$
and a regularized $R'(0)$ in QCD. 

In contrast, the term in the P-wave decay rate that is proportional to
${\cal H}_8$ is absent in the color-singlet model.  ${\cal H}_8$ is the
probability to find a $Q\bar{Q}g$ component in P-wave quarkonium, with
the $Q\bar Q$ pair in a relative S-wave, color-octet state.  As such, it
corresponds to a true field-theoretic effect of QCD that is absent in
any potential model of quarkonium.  

\narrowtext 
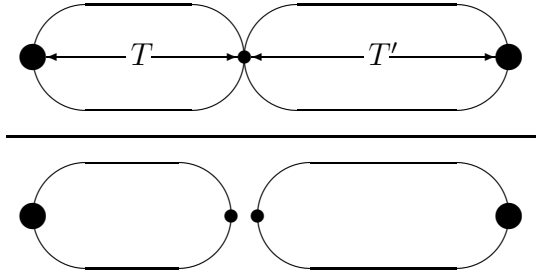
\begin{figure}[htb]
\centerline{ 
\begin{picture}(200,125)
\put(50,95){\oval(80,40)}
\put(140,95){\oval(100,40)}
\put(0,65){\line(1,0){200}}
\put(47.5,35){\oval(75,40)}
\put(142.5,35){\oval(95,40)}
\put(10,95){\circle*{10}}
\put(190,95){\circle*{10}}
\put(10,35){\circle*{10}}
\put(190,35){\circle*{10}}
\put(90,95){\circle*{5}}
\put(85,35){\circle*{5}}
\put(95,35){\circle*{5}}
\put(45,95){\vector(-1,0){30}}
\put(55,95){\vector(1,0){32.5}}
\put(135,95){\vector(-1,0){42.5}}
\put(145,95){\vector(1,0){40}}
\put(47,92){$T$}
\put(137,92){$T'$}
\end{picture}}
\caption{Lattice calculation of matrix element of 4-fermion operator.
The large discs represent the sources and sinks; the smaller discs
represent the 4-fermion and point source operators. The lines are the 
nonrelativistic quark propagators.} 
\label{fig:matrix}
\end{figure}
\widetext
On the lattice, the long-distance matrix elements are obtained from the 
graphs of Fig.~\ref{fig:matrix}.  The upper and lower graphs yield
quantities that fall as $ \exp [-E(|T|+|T'|)]$.  The matrix element is given
by the limit as $T,T'\rightarrow \infty$ of the ratio of 
the upper graph to the lower graph, 
with the same choice of sources and sinks in both graphs, times the coefficient
of the exponential fall off for the point-point quarkonium propagator. 
We used noisy-point and noisy-gaussian sources and generated retarded
and advanced quark propagators from each time slice. We chose
the Coulomb gauge for the field configurations.  This choice made
implementation of extended sources simpler and allowed us to replace
covariant derivatives with normal derivatives, with errors of relative
order $v^2$. We calculated heavy-quark propagators $G({\bf x},t)$ on the
lattice, using the nonrelativistic formulation of Lepage {\it et al.}
\cite{lepage}, with an evolution equation that is valid to the lowest
non-trivial order in $v^2$: 
\begin{eqnarray}
G({\bf x},t+1)=(1-H_0/2n)^n U_{{\bf x},t}^{\dag} (1&-&H_0/2n)^n G({\bf x},t)
                                                         \nonumber \\
                                          &+&\delta_{\bf x,0}\delta_{t+1,0},
\end{eqnarray}
with $G({\bf x},t)=0$ for $t < 0$, and $H_0 = -\Delta^{(2)}/2 M_0 -
h_0$. $\Delta^{(2)}$ is the gauge-covariant discrete Laplacian, $M_0$
the bare heavy-quark mass, and $h_0=3(1-u_0)/M_0$, with
$u_0=\langle\frac{1}{3}U_{plaqette}\rangle^\frac{1}{4}$. We chose $n=2$.
Since ${\cal F}_1/M_Q^2$ is suppressed by ${\cal O}(v^2)$ relative to
${\cal G}_1$, it is of the same order as terms that we have neglected in
the computation of ${\cal G}_1$. The main justification for its
calculation is that there are decays such as $^3S_1 \rightarrow \hbox{Light
Hadrons}$, $^3S_1 \rightarrow \gamma + \hbox{Light Hadrons}$, and $^3S_1
\rightarrow 3\gamma$, for which the coefficient of ${\cal F}_1/M_Q^2$ is
approximately $-5$ times that of ${\cal G}_1$ \cite{bbl3,bbl}.  In these
cases, the contributions of terms proportional to ${\cal F}_1$ could be
important. 

In our lattice calculations we used 149 quenched gauge-field
configurations on a $16^3 \times 32$ lattice with $6/g^2=6.0$ for
bottomonium and 158 configurations on an $16^3 \times 32$ lattice with
$6/g^2=5.7$ for bottomonium and charmonium.  For bottomonium we took
$M_0=1.5$ at $6/g^2=6.0$ and $M_0=2.7$ at $6/g^2=5.7$. For charmonium
at $6/g^2=5.7$ we took $M_0=0.69$. These values correspond to those used 
by the NRQCD collaboration \cite{nrqcd}. (Note that their mass
definitions are $u_0$ times ours.) The values of $u_0$ that we used are
0.87778701 at $6/g^2=6.0$ and  0.8608261760 at $6/g^2=5.7$. Except where
we explicitly state otherwise, all quantities in this paper are in
lattice units. To convert to physical units, we use inverse lattice
spacings $a^{-1}=2.4$~GeV for bottomonium at $6/g^2=6.0$,
$a^{-1}=1.37$~GeV for bottomonium at $6/g^2=5.7$, and $a^{-1}=1.23$~GeV
for charmonium at $6/g^2=5.7$.  These are the values obtained by the
NRQCD collaboration. Our error estimates do not include the errors in
these quantities. 

NRQCD predicts that \cite{bbl}
\begin{eqnarray}
{\cal G}_1/|\langle^1S_0|\psi^{\dag}\chi|0\rangle|^2 & = & (1+{\cal O}(v^4)) \\
{\cal H}_1/|\langle^1P_1|\psi^{\dag}\frac{-i}{2}\stackrel{\leftrightarrow}
                        {\bf D}\chi|0\rangle|^2 & = & (1+{\cal O}(v^4)),
\end{eqnarray}
where the vacuum-saturation approximation amounts, in this case, to
ignoring the ${\cal O}(v^4)$ term. For bottomonium at $6/g^2=6.0$ we
measured the $v^4$ term for ${\cal G}_1$ to be $1.3(1) \times 10^{-3}$.
For charmonium at $6/g^2=5.7$ this term is approximately $1\%$. For
${\cal H}_1$, these ${\cal O}(v^4)$ terms, while larger than those for
${\cal G}_1$, are still quite small. Thus, the vacuum saturation
approximation is even better than one would expect. We will therefore
use the vacuum-saturation values for ${\cal G}_1$, ${\cal H}_1$, and
${\cal F}_1$ in the discussions to follow. The lattice quantities ${\cal
G}_{1L}$ and ${\cal H}_{1L}$ are then given by the coefficients of the
exponentials in the S- and P-wave quarkonium propagators, respectively. 

\narrowtext
\begin{table}[htb]

\begin{tabular}{|@{\hspace{0.2cm}}l@{\hspace{0.2cm}}%
|c@{\hspace{1.2cm}}|c@{\hspace{1.2cm}}|c@{\hspace{1.2cm}}|}
              & charmonium & \multicolumn{2}{c@{\hspace{8mm}}|}{bottomonium} \\
\hline
$6/g^2$                  & 5.7        & 5.7            & 6.0             \\
\hline
${\cal G}_{1L}$                      & 0.1317(2)(12) & 0.9156(9)(65) 
                                                     & 0.1489(5)(12) \\ 
${\cal F}_{1L}(non)/{\cal G}_{1L}$   & 1.2543(7)     & 2.7456(8) 
                                                     & 1.3135(8)     \\
${\cal F}_{1L}(cov)/{\cal G}_{1L}$   & 0.5950(5)     & 2.1547(7)
                                                     & 0.8522(5)     \\
${\cal F}_{1L}(non_2)/{\cal G}_{1L}$ & 0.7534(4)     & 1.2205(2) 
                                                     & 0.7775(5)     \\  
${\cal F}_{1L}(cov_2)/{\cal G}_{1L}$ & 0.5201(3)     & 1.1111(2) 
                                                     & 0.6659(3)     \\
${\cal H}_{1L}$                      & 0.0208(2)(20) &   -----     
                                                     & 0.0145(6)(20) \\
${\cal H}_{8L}/{\cal H}_{1L}$        & 0.034(2)(8)   &   -----     
                                                     & 0.0152(3)(20) \\
\end{tabular}
\caption{Lattice decay matrix elements expressed in lattice
units ($a=1$). Note that P-wave bottomonium matrix elements have yet to
be calculated at $6/g^2=5.7$.} 
\label{tab:lattice}
\end{table}
\widetext
A summary of our results for the lattice matrix elements defined in
(\ref{matrix-elements}) is presented in Table~\ref{tab:lattice}. 
When a second error has been included, it is an estimate of the
systematic errors associated with the parametrization of the fitting
functions and with the contamination from higher states for propagators
in which the separation between source and sink is too small.

To the order in $v^2$ in which we are working, our lattice matrix elements are
related to their continuum counterparts by 
\begin{mathletters}
\label{eqn:s-wave}
\begin{eqnarray}
{\cal G}_{1L} &=& (1+\epsilon) {\cal G}_1, \\
{\cal F}_{1L} &=& (1+\gamma ) {\cal F}_1 + \phi {\cal 
G}_1,\label{eqn:s-wave-unit-op}
\end{eqnarray}
\end{mathletters}
and
\begin{mathletters}
\label{eqn:p-wave}
\begin{eqnarray}
{\cal H}_{1L} &=& (1+\iota ) {\cal H}_1 + \kappa {\cal H}_8, \\
{\cal H}_{8L} &=& (1+\eta) {\cal H}_8 + \zeta {\cal H}_1,
\end{eqnarray}
\end{mathletters}%
where the subscript $L$ indicates the lattice quantity. The coefficients
$\epsilon$, $\gamma$, $\phi$, $\iota$, $\eta$ and $\zeta$ are of order
$\alpha_s$; $\kappa$ is of order $\alpha_s^2$. We have calculated these
coefficients through order $\alpha_s$ (one loop) in tadpole-improved
perturbation theory \cite{lm}. Our values for these coefficients, for
$\overline{MS}$~regularization of the continuum matrix elements, are
given in Table~\ref{tab:coeff}. The accuracy of the coefficients of
$\alpha_s$ in this table is estimated to be better than 1\%. In
computing $\zeta$, we have taken the factorization scale to be 1.3~GeV
for charmonium and 4.3~GeV for bottomonium. These values correspond,
approximately, to the $\overline {MS}$ heavy-quark masses. Note that
$\phi$ and $\kappa$, in physical units, have dimensions of
$(\hbox{mass})^2$, and $\zeta$ has dimensions of $1/\hbox{mass}^2$,
whereas the other coefficients are dimensionless.  If we render $\phi$,
$\kappa$, and $\zeta$ dimensionless by dividing ${\cal F}_1$ and $\phi$
by $M_Q^2$, by dividing ${\cal H}_1$ and $\kappa$ by $M_Q^2$, and by
multiplying $\zeta$ by $M_Q^2$, respectively, then none of the
coefficients of $\alpha_s$ is exceptionally large. Hence, the use of
low-order perturbation theory appears to be reasonable. 
\narrowtext
\begin{table}[htb]
\begin{tabular}{|@{\hspace{0.2cm}}l@{\hspace{0.2cm}}%
|c@{\hspace{1.4cm}}|c@{\hspace{1.4cm}}|c@{\hspace{1.4cm}}|}
         & charmonium & \multicolumn{2}{c@{\hspace{1cm}}|}{bottomonium} \\
\hline
$6/g^2$         & 5.7        & 5.7            & 6.0             \\
\hline
$\epsilon$      & -0.7326 $\alpha_s$ & 0.2983   $\alpha_s$ 
                & -0.4877 $\alpha_s$  \\
$\gamma(non)$   & -0.02578 $\alpha_s$ & -1.248   $\alpha_s$ 
                & -0.9117 $\alpha_s$  \\
$\gamma(cov)$   & -2.860   $\alpha_s$ & -2.192   $\alpha_s$ 
                & -2.560  $\alpha_s$  \\
$\gamma(non_2)$ & -0.2774 $\alpha_s$ & -1.096   $\alpha_s$ 
                & -0.9236 $\alpha_s$  \\
$\phi(non)$     & 1.486   $\alpha_s$ & 10.90    $\alpha_s$ 
                & 4.418   $\alpha_s$  \\
$\phi(cov)$     & 0.3928  $\alpha_s$ & 9.808    $\alpha_s$ 
                & 3.325    $\alpha_s$  \\
$\phi(non_2)$   & 1.004  $\alpha_s$ & 6.096   $\alpha_s$ 
                & 2.863  $\alpha_s$  \\
$\iota$         & -0.7603 $\alpha_s$ & -1.852   $\alpha_s$ 
                & -1.191  $\alpha_s$  \\
$\eta$          & 0.09157 $\alpha_s$ & -0.03728 $\alpha_s$ 
                & 0.06096 $\alpha_s$  \\
$\zeta$         & -0.1785 $\alpha_s$ & -0.006011 $\alpha_s$ 
                & -0.01862 $\alpha_s$
\end{tabular}
\caption{Coefficients relating lattice and continuum matrix elements. The
arguments of $\gamma$ and $\phi$ correspond to different lattice
representations of ${\cal F}_1$. $cov$ is a tadpole-improved
\protect\cite{lepage} naive discretization of the gauge covariant continuum
operator; $non$ is the simple, gauge-noncovariant, finite-difference 
operator in coulomb gauge; the subscript $2$ indicates a
difference operator with spacings of 2 lattice units.} 
\label{tab:coeff}
\end{table}
\widetext

For those coefficients that arise from a positive integrand, the method
of Lepage and Mackenzie \cite{lm} yields an optimal scale for $\alpha_s$
that is close to $1/a$.  Thus, we choose $\alpha_s=\alpha_V (1/a)=0.3552$
at $6/g^2=5.7$ and $0.2467$ at $6/g^2=6.0$. 

Substituting the numerical values from Tables~\ref{tab:lattice} and
\ref{tab:coeff} 
into (\ref{eqn:s-wave}) and (\ref{eqn:p-wave}), we obtain the results 
shown in Table~\ref{tab:cont}.
In Table~\ref{tab:cont}, the first and second errors in the lattice
results are from the statistical and systematic errors in
Table~\ref{tab:lattice}. The third error is an estimate of the
systematic error that arises from the neglect of terms of higher order
in $\alpha_s$ in the coefficients of Table~\ref{tab:coeff}. It is
obtained by taking the uncertainty in the coefficients to be either
$\alpha_s^2$ times the zeroth order term (if any) or $\alpha_s$ times
the magnitude of the first order term, whichever is the larger. In the
case of ${\cal F}_1/{\cal G}_1$, the uncertainty is large, so we have
presented our results as a range of values. 
\narrowtext
\begin{table}[htb]
\begin{tabular}{|@{\hspace{0.2cm}}l@{\hspace{0.2cm}}%
|c@{\hspace{1.2cm}}|c@{\hspace{1.2cm}}|c@{\hspace{1.2cm}}|}
             & \multicolumn{2}{c@{\hspace{1.6cm}}|}{lattice} & experiment  \\
\hline
                         & lattice units       & physical units &          \\
\hline
\multicolumn{2}{|l|}{charmonium $6/g^2=5.7$}   &                &          \\ 
${\cal G}_1$                  & 0.1780(3)(16)$^{+366}_{-259}$
                              & 0.3312(6)(30)$^{+681}_{-483}$~GeV$^3$
                              & 0.36(3)~GeV$^3$                            \\
${\cal F}_1 / {\cal G}_1$       & 0.05 --- 0.54 
                              & 0.07 --- 0.82~GeV$^2$
                              & 0.057~GeV$^2$                              \\
${\cal H}_1$                  & 0.0285(2)(27)$^{+60}_{-42}$
                              & 0.0802(6)(77)$^{+167}_{-118}$~GeV$^5$
                              & 0.077(19)(28)                              \\
        &          &          & $\;\;\;\;\;$~GeV$^5$                       \\
${\cal H}_8 / {\cal H}_1$     & 0.086(1)(6)$^{+42}_{-32}$
                              & 0.057(1)(4)$^{+27}_{-21}$~GeV$^{-2}$
                              & 0.095(31)(34)                              \\
        &          &          & $\;\;\;\;\;$~GeV$^{-2}$                    \\
\hline
\multicolumn{2}{|l|}{bottomonium $6/g^2=5.7$}    &                &        \\
${\cal G}_1$                  & 0.8279(8)(59)$^{+1066}_{-848}$
                              & 2.129(2)(15)$^{+274}_{-218}$~GeV$^3$
                              & 3.55(8)~GeV$^3$                            \\
${\cal F}_1 / {\cal G}_1$     & -3.7 --- 0.2
                              & -6.9 --- 0.4~GeV$^2$
                              &     -----                                  \\
\hline
\multicolumn{2}{|l|}{bottomonium $6/g^2=6.0$}    &                &        \\
${\cal G}_1$                  & 0.1692(6)(14)$^{+126}_{-110}$
                              & 2.340(8)(19)$^{+173}_{-151}$~GeV$^3$
                              & 3.55(8)~GeV$^3$                            \\
${\cal F}_1 / {\cal G}_1$     & -0.34 --- 0.28 
                              & -2.0 --- 1.6~GeV$^2$
                              &     -----                                  \\
${\cal H}_1$                  & 0.0205(9)(28)$^{+23}_{-19}$
                              & 1.63(7)(23)$^{+19}_{-15}$~GeV$^5$
                              &     -----                                  \\
${\cal H}_8 / {\cal H}_1$     & 0.0151(2)(14)$^{+33}_{-29}$
                              & 0.00262(3)(24)$^{+57}_{-51}$GeV$^{-2}$
                              &     -----                                 
\end{tabular}
\caption{Continuum $\overline{MS}$ decay matrix elements from our lattice
calculations, compared with those extracted from experimental decay rates,
where available.}
\label{tab:cont}
\end{table}
\widetext

For purposes of comparison, we have also shown in Table~\ref{tab:cont}
the experimental (phenomenological) results, where available, for the
matrix elements that we have computed.  The phenomenological results for
${\cal G}_1$ were extracted from the measured decay rates for $J/\psi
\rightarrow e^+ e^-$, $\eta_c \rightarrow \gamma\gamma$ and $\Upsilon
\rightarrow e^+ e^-$ \cite{rpp}, using the expressions in
Ref.~\cite{bbl}. Values for ${\cal F}_1/{\cal G}_1$ for $J/\psi$ are
those of Ko, Lee and Song \cite{kls}. The results for ${\cal H}_1$ and
${\cal H}_8/{\cal H}_1$ for $\chi_c$ are from \cite{bbl2}; the first
error is experimental, the second theoretical. For P-wave bottomonium,
there is as yet no published data on decays into light hadrons, photons,
and/or leptons. The extraction of phenomenological matrix elements from
the experimental data requires values for the heavy-quark masses. Our
choices correspond to pole masses of 5.0~GeV for the b~quark, the result
obtained by the NRQCD collaboration \cite{nrqcd}, and 1.5~GeV for the
c~quark \cite{eq}. We also
require values for $\alpha_s$ and $\alpha$ in order to evaluate the
partonic decay rates. For these we used $\alpha_s(M_c)=0.243$,
$\alpha_s(M_b)=0.179$, $\alpha(M_c)=1/133.3$ and $\alpha(M_b)=1/132$. 

In the above analysis, we have not taken into account the errors due to
the omission of terms of higher order in $v^2$. These could be as large
as 10\% for bottomonium and 30\% for charmonium. For ${\cal G}_{1L}$,
the NRQCD collaboration has published results that are accurate to the
next-to-leading order in $v^2$. Since these higher-order results
distinguish the singlet and triplet states, we compare our results with
a weighted average of their results. For charmonium at $6/g^2=5.7$ they
obtain ${\cal G}_{1L}=0.133(4)$, and for bottomonium at $6/g^2=6.0$ they
obtain ${\cal G}_{1L}=0.144(4)$, in good agreement with the results of
Table~\ref{tab:lattice}. For these matrix elements, as with masses, most
of the effect of contributions of higher order in $v^2$ is to split the
results for the singlet and triplet states without shifting the weighted
average. 

There are some additional sources of error that we have not included in
Table~\ref{tab:cont}. One of these is the uncertainty in the physical
value of $a^{-1}$ . Using the results of the NRQCD collaboration, we
find that the uncertainties in the values of the matrix elements from
this source are 7\% for ${\cal G}_1$ and 13\% for ${\cal H}_1$ in
charmonium and 13\% for ${\cal G}_1$ and 23\% for ${\cal H}_1$ in
bottomonium. As we have already mentioned, extraction of the
phenomenological matrix elements requires knowledge of the heavy-quark
mass. The NRQCD collaboration quotes an error of 4\% for the $b$-quark mass,
which introduces an 8\% error in ${\cal G}_1$ and a 16\% error in ${\cal
H}_1$. For the $c$-quark mass we have no good error estimate. Further
sources of uncertainty are the QCD radiative corrections to parton-level
decay rates. Estimates of these uncertainties have been included in the
phenomenological values of ${\cal H}_1$ and ${\cal H}_8$ for charmonium
that are reported in Ref.~\cite{bbl2}. Finally, there are the errors
that arise from using quenched (rather than full) QCD, for which we have
no estimates. 

Let us now discuss our results. For charmonium ${\cal G}_1$, ${\cal
H}_1$, and ${\cal H}_8/{\cal H}_1$ are in agreement with experiment,
although both the lattice and experimental results have sizeable
errors. We note that we would not have found this agreement had we failed to
include the perturbative corrections that relate the lattice matrix
elements to the continuum ones. The quantity ${\cal F}_1/{\cal G}_1$ is
poorly determined, for both charmonium and bottomonium, owing to the
mixing of ${\cal F}_1$ with ${\cal G}_1$ in (\ref{eqn:s-wave}).
Because ${\cal F}_1/(M^2{\cal G}_1)$ is of order $v^2\ll 1$, a
coefficient $\phi/M^2$ of order $\alpha_s$ yields a large mixing, and
any uncertainties in $\phi/M^2$ are amplified in ${\cal F}_1/(M^2{\cal
G}_1)$. We do learn, though, that ${\cal F}_1/(M^2{\cal G}_1)$ is no
larger than ${\cal O}(v^2)$, in agreement with the NRQCD scaling rules
\cite{lepage,bbl}. For charmonium, ${\cal F}_1/{\cal G}_1$ is probably
positive, while for bottomonium a negative value is preferred. In the
case of bottomonium, the lattice result for ${\cal G}_1$ is 35 --- 40\%
below the experimental value, although there is good agreement between
the $6/g^2=5.7$ and $6/g^2=6.0$ predictions. At least part of this
discrepancy, which was first noted by the NRQCD collaboration, is due to
the quenched approximation \cite{nrqcd,gpl}.  Our results for the P-wave
matrix elements for bottomonium can be translated into predictions for
bottomonium decay rates \cite{bbl2}.  In the P-wave case, these should
lead to significant new tests of the theory as the relevant experimental
data become available. 

As is clear from Table~\ref{tab:cont}, the largest uncertainties in the
lattice matrix elements (aside from those due to quenching) come from
neglecting higher-order corrections to the coefficients of
Table~\ref{tab:coeff}. This suggests that a useful strategy might be to
use lattice methods \cite{sachrajda} to compute the coefficients beyond
leading order.  In addition, one might consider the use of alternatives
to the $\overline {MS}$ regularization of the continuum matrix elements
so as to avoid renormalon ambiguities in the matrix elements and
short-distance coefficients\cite{sachrajda}. Note that, to the extent
that we can replace ${\cal H}_{1L}$ and ${\cal H}_1$ by their vacuum
saturation approximations, ${\cal H}_1$ and ${\cal H}_8$ are free of
renormalon ambiguities.  Of course, these ambiguities cancel in physical
quantities if one works consistently to a given order in $\alpha_s$ in
both the lattice-to-continuum coefficients and the short-distance
coefficients. 

It is an interesting fact that, for decays of P-wave states in both charmonium
and bottomonium, the ratio of the octet matrix element to the singlet matrix
element is in reasonable agreement with a crude phenomenology.  This
phenomenology is based on solving the one-loop evolution equation for ${\cal
H}_8$ (Ref.~\cite{bbl}) and assuming that ${\cal H}_8$ vanishes below a scale
$M_Q v$. Since the one-loop evolution contribution to the decay matrix element
${\cal H}_8$ is the same as that for the corresponding production matrix
element ${\cal H}'_8$, this simple phenomenology suggests that ${\cal H}'_8$ is
approximately equal to ${\cal H}_8$. For charmonium production, ${\cal H}'_8$
has been extracted from CDF data \cite{cl} and  can also be deduced from recent
CLEO data \cite{cleo}. Dividing the CDF and CLEO values of ${\cal H}'_8$ by the
phenomenological value of ${\cal H}_1$ given in Table~\ref{tab:cont}, we obtain
0.042(19)GeV$^{-2}$ and 0.046(28)GeV$^{-2}$ respectively, which agree, within
errors, with our lattice result for ${\cal H}_8/{\cal H}_1$. 

We wish to thank G.~Peter~Lepage, John Sloan and Christine Davies for
informative discussions and for access to some of their unpublished results. We
also thank G.~Peter~Lepage for carrying out simulations to check our S-wave
results. Our calculations were performed on the CRAY C-90 at NERSC, whose
resources were made available to us through the Energy Research Division of the
U.~S. Department of Energy. This work was supported by the U.S. Department of 
Energy, Division of High Energy Physics, Contract W-31-109-ENG-38.
S.~K. is supported by KOSEF through CTP.
A preliminary version of these results was
presented at LATTICE~'94 \cite{bks}.

\end{document}